\documentclass[3p,times]{elsarticle}

\usepackage{ecrc}

\volume{00}

\firstpage{1}

\journalname{Nuclear Physics A}

\runauth{W.\ A.\ Horowitz}

\jid{nupha}

\usepackage{amssymb}

 \biboptions{sort&compress,numbers,comma,square}

\usepackage{color}
\definecolor{darkgreen}{rgb}{0,.7,0}
\definecolor{linkblue}{rgb}{0.,0.,0.9333}
\usepackage{hyperref} 
\hypersetup{
    pdffitwindow=true,      
    pdfauthor={W. A. Horowitz},     
    pdfnewwindow=true,      
    bookmarksnumbered=true,
    colorlinks=true,       
    linkcolor=red,          
    citecolor=darkgreen,        
    filecolor=magenta,      
    urlcolor=cyan,           
    menucolor=blue,
}

\newcommand{\be}{\begin{equation}}
\newcommand{\ee}{\end{equation}}

\newcommand{\fig}[1]{Fig.~\ref{#1}}

\begin{document}

\begin{frontmatter}



\dochead{}

\title{Weakness or Strength in the Golden Years of RHIC and LHC?}

\author{W.\ A.\ Horowitz}

\address{Department of Physics, University of Cape Town, Private Bag X3, Rondebosch 7701, South Africa}

\begin{abstract}
Recent LHC data suggest that perturbative QCD provides a qualitatively consistent picture of jet quenching.  Constrained to RHIC $\pi^0$ suppression, zero parameter WHDG energy loss predictions agree quantitatively with the charged hadron $v_2$ and $D$ meson $R_{AA}$ measured at LHC and qualitatively with the $h^\pm$ $R_{AA}$.  On the other hand, RHIC-constrained LHC predictions from fully strongly-coupled AdS/CFT qualitatively oversuppress $D$ mesons compared to data; light meson predictions are on less firm theoretical ground but also suggest oversuppression.  More detailed data from heavy, especially $B$, mesons will continue to help clarify our picture of the physics of the quark-gluon plasma. Since the approach of pQCD predictions to LHC data occurs at momenta $\gtrsim 15$ GeV/c, a robust consistency check between pQCD and both RHIC and LHC data requires RHIC jet measurements.
\end{abstract}


\end{frontmatter}

\section{Introduction}
Jet quenching calculations compared to data provide the most direct and detailed probe of the relevant degrees of freedom of the quark-gluon plasma created in relativistic heavy ion collisions at RHIC and LHC \cite{Armesto:2011ht}.  While early comparisons of pQCD-based energy loss calculations to the measured light hadron suppression at RHIC were successful \cite{Vitev:2002pf}, those made to heavy quark suppression and the azimuthal anisotropy of the data showed a quantitative disagreement \cite{Wicks:2005gt,Adare:2010sp}.  Subsequent to the success of strongly-coupled AdS/CFT techniques in describing the very small measured viscosity to entropy ratio, $\eta/s\sim0.1$, extracted from hydrodynamic studies at RHIC and LHC \cite{Kovtun:2004de,Schenke:2010rr}, these tools were used \cite{Gubser:2006bz,Herzog:2006gh} to quantitatively describe the suppression of the remnants of high-$p_T$ heavy quarks at RHIC \cite{Akamatsu:2008ge,Horowitz:2010dm}.  The exciting new data from LHC begs for comparison with theoretical calculations.  

In this paper we will use the WHDG energy loss model \cite{Wicks:2005gt} to compare perturbative QCD (pQCD) calculations to data.  In this model, radiative energy loss from the pQCD-based DGLV opacity expansion derivation is incoherently convolved with pQCD-based collisional energy loss.  Due to the finite kinematics relevant in RHIC and LHC collisions, the magnitude of the collisional loss is of the same order as the radiative loss and is crucially important for consistently describing both the heavy and light meson suppression \cite{Wicks:2005gt}.  While there are large systematic theoretical uncertainties in the inelastic energy loss due to the collinear approximation \cite{Horowitz:2009eb}, these uncertainties are small for predictions when the medium density is constrained by data \cite{Adare:2008cg,Horowitz:2011gd}, like those presented in these proceedings.  AdS/CFT energy loss predictions here are implemented \cite{Horowitz:2007su} based on the heavy quark drag derivations \cite{Gubser:2006bz,Herzog:2006gh}.

\section{Energy Loss Model Results}

\begin{figure}[!hbp]
\centering
$\begin{array}{ccc}
\includegraphics[width=2.5in]{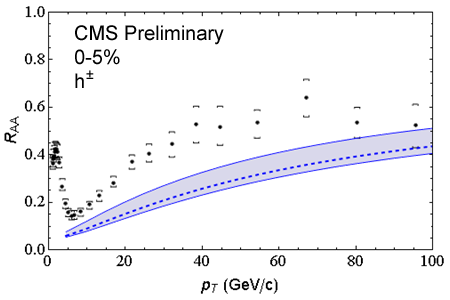} & \hspace{.02in} & 
\includegraphics[width=2.6in]{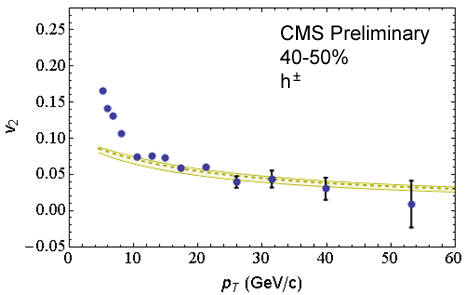}
\end{array}$
\vspace{-.3in}
\begin{flushleft}
$\begin{array}{ccc}
\hspace{.3in} \mbox{(a)} & \hspace{2.35in} & \mbox{(b)}
\end{array}$
\end{flushleft}
\centering
\includegraphics[width=2.5in]{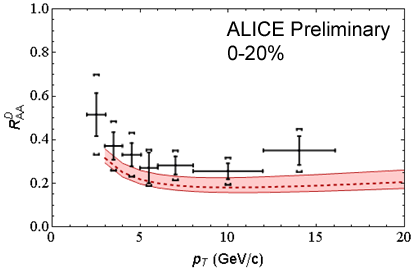}
\vspace{-.3in}
\begin{flushleft}
$\begin{array}{c}
\hspace{1.8in} \mbox{(c)}
\end{array}$
\end{flushleft}
\vspace{-.1in}
\caption{
\label{WHDGlhc}
The plotted pQCD calculations are based on the WHDG model of energy loss, which assumes partons are weakly coupled to a weakly coupled quark-gluon plasma  \cite{Wicks:2005gt} and are constrained to RHIC $\pi^0$ $R_{AA}$ data \cite{Adare:2008cg,Horowitz:2011gd}. (a) $R_{AA}(p_T)$ of pions at mid-rapidity and in most central collisions compared to LHC charged hadron data \cite{CMS:2012aa}. (b) $v_{2}(p_T)$ of pions at mid-rapidity and at 40-50\% centrality compared to LHC charged hadron data \cite{Chatrchyan:2012xq}. (c) $R_{AA}(p_T)$ for $D$ mesons at mid-rapidity and 0-20\% centrality compared to LHC data \cite{ALICE:2012ab}.}
\end{figure}

The charged particle suppression was the first jet quenching measurement to emerge from LHC \cite{Aamodt:2010jd,CMS:2012aa}.  Compared to zero parameter predictions constrained to RHIC data \cite{Adare:2008cg,Horowitz:2011gd}, the LHC data appeared surprisingly transparent, \fig{WHDGlhc} (a).  Here zero parameter predictions means keeping all theoretical input values fixed except for the QGP medium density, which is assumed to scale linearly with the observed charged particle multiplicity.  The LHC charged particle $R_{AA}$ taken together with the more recent LHC $D$ meson suppression and $v_2$ of the charged hadron data, along with the RHIC $\pi^0$ $R_{AA}$ and $v_2$ and non-photonic electron $R_{AA}$, the zero parameter WHDG predictions manage a surprisingly good global description of the data, on the order of $\pm50\%$; see \fig{WHDGlhc} (b) and (c).  This level of accuracy is similar to the next-to-leading order pQCD predictions of particle production in hadronic collisions \cite{Adare:2010de,CMS:2012aa}.  The NLO pQCD production calculation is much cleaner theoretically, and we must therefore consider its level of discrepancy as a good estimate of the lower bound of the discrepancy we should expect from an honest comparison between QCD calculations and data.  In particular, the level of agreement between the WHDG calculations and data is in a sense a pleasant surprise given that WHDG is only a leading order computation.

\begin{figure}[!htbp]
\centering
$\begin{array}{ccc}
\includegraphics[width=1.74in]{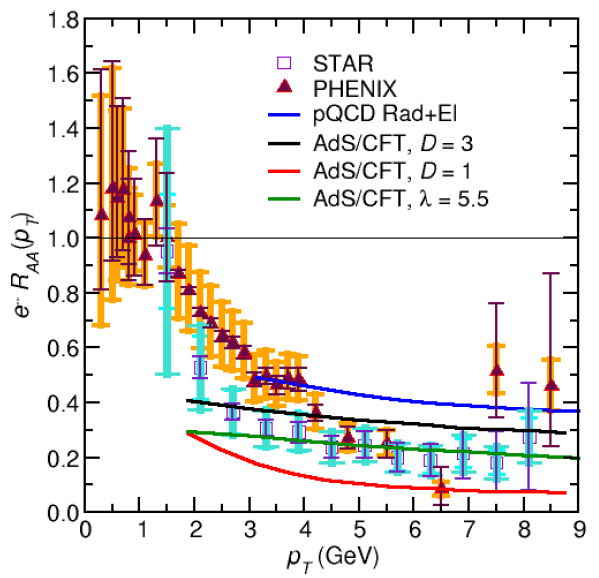} & \hspace{.3in} &
\includegraphics[width=2.5in]{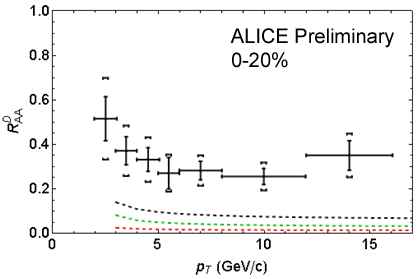}
\end{array}$
\vspace{-.3in}
\begin{flushleft}
$\begin{array}{ccc}
\hspace{.5in} \mbox{(a)} & \hspace{2.2in} & \mbox{(b)}
\end{array}$
\end{flushleft}
\vspace{-.1in}
\caption{
\label{ads}
(a) Comparison of the measured non-photonic electron suppression at RHIC \cite{Adare:2006nq,Abelev:2006db} and predictions from an AdS/CFT drag model \cite{Horowitz:2007su,Horowitz:2010dm}. (b) Comparison of measured $D$ meson suppression at LHC \cite{ALICE:2012ab} and predictions from the same AdS/CFT drag model \cite{Horowitz:2011wm}.}
\end{figure}

Just like the pQCD calculations, one may make simultaneous comparisons between results predicted by AdS/CFT and RHIC and LHC data.  Since the light quark energy loss derivation is under less theoretical control, we compare only to data related to heavy quark suppression.  It is also nontrivial to translate the relevant QCD parameters into those used in the AdS/CFT calculation; a reasonable exploration of the possibilities is represented by the three curves shown in \fig{ads} (a), which quantitatively describe the suppression of non-photonic electrons at RHIC.  Making the same zero parameter extrapolation to LHC as was done for the pQCD calculation results in a $D$ meson suppression which is much greater than that observed by the data, \fig{ads} (b).  In fact, the disagreement with the central data values is qualitative as opposed to quantitative as in the perturbative case; however, the current experimental uncertainties are large and do not seem to completely preclude the possibility of a non-perturbative description of the data.  

Despite significant efforts \cite{Gubser:2008as,Chesler:2008uy,Ficnar:2012nu}, AdS/CFT estimates for light quark and gluon energy loss are qualitative at best.  Nevertheless, naive application of AdS/CFT techniques robustly suggest that even very large momentum quarks ($\sim 100$ GeV/c) strongly coupled to a strongly coupled medium of constant temperature thermalize extremely rapidly, $\sim2.7$ fm.  Recent work \cite{MoradHorowitz} has found that when Bjorken expansion is included and the temperature decreases with time the thermalization distance increases to $\sim4.1$ fm.  Taken seriously, it is difficult to imagine that a relatively sophisticated estimate of the suppression would be consistent with data.  

\begin{figure}[!hbp]
\centering
$\begin{array}{ccc}
\includegraphics[width=2.5in]{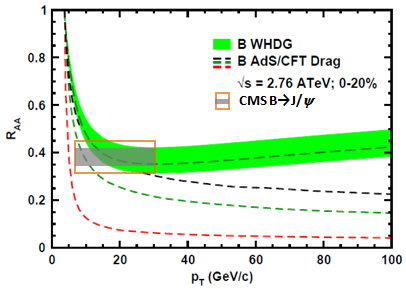} & \hspace{.3in} &
\includegraphics[width=2.5in]{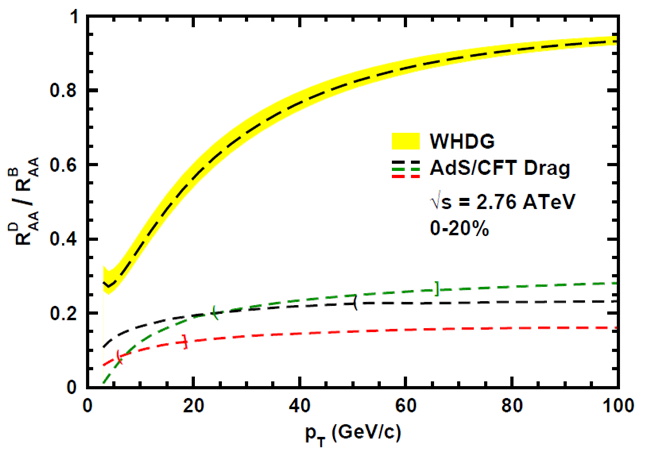}
\end{array}$
\vspace{-.3in}
\begin{flushleft}
$\begin{array}{ccc}
\hspace{.5in} \mbox{(a)} & \hspace{2.2in} & \mbox{(b)}
\end{array}$
\end{flushleft}
\vspace{-.1in}
\caption{
\label{adsvspqcd}
(a) Comparison of the measured non-prompt $J/\psi$ mesons, the decay products of $B$ mesons, as measured by CMS at LHC \cite{Chatrchyan:2012np} to both AdS/CFT \cite{Horowitz:2007su,Horowitz:2011wm} and pQCD predictions \cite{Wicks:2005gt,Horowitz:2011gd,Horowitz:2011wm}. (b) The double ratio of $D$ meson $R_{AA}$ to $B$ meson $R_{AA}$ as a function of $p_T$ from WHDG \cite{Wicks:2005gt} and an AdS/CFT drag model \cite{Horowitz:2007su,Horowitz:2011wm}.}
\end{figure}

There are already some tantalizing results on $B$ meson suppression from CMS.  In \fig{adsvspqcd} (a) we see that both AdS/CFT and WHDG calculations are consistent with the current experimental results.  Future measurements will hopefully be more distinguishing.  In particular, the generic predictions of the mass and momentum dependence of AdS/CFT and pQCD are strikingly different; we show this difference in \fig{adsvspqcd} (b).  Generically, the perturbative calculations ``lose track'' of the mass of the parent parton as the momentum scale increases; on the other hand, to leading order, the AdS calculations do not.  The suppression of open charm and bottom production in p + A collisions will provide another interesting test of the AdS/CFT drag formalism \cite{Horowitz:2009pw}.

\section{Conclusions}
Despite the success of AdS/CFT in readily describing the near-perfect fluidity of QGP from leading order results, we find that high-$p_T$ observables are best described assuming a weakly coupled probe interacting with a weakly coupled medium.  AdS/CFT calculations of the energy loss of a probe strongly coupled to a strongly coupled medium lead to an overprediction of the suppression compared to data.  Perhaps next-to-leading order effects, both from the pQCD side and from the AdS/CFT side will shed light on this discrepancy.

It is worth emphasizing that---lacking a theoretical handle on NLO effects---one can see the apparent onset of the applicability of pQCD techniques by comparing the calculations to data in \fig{WHDGlhc} at $\gtrsim$ 15 - 20 GeV/c; the discrepancy below these momenta seem to imply that non-perturbative effects are large.  Due to limited statistics, current and future single particle observables at RHIC in the momentum range where pQCD techniques describe the data well are limited to only light hadron $R_{AA}$.  An experimental test of the consistent, coherent picture of pQCD energy loss techniques as applied in heavy ion collisions for partons of momenta $\gtrsim$ 15 - 20 GeV/c would require measurements of more differential observables and also heavy mesons; these measurements, in turn, for purely statistical reasons, require the measurement of jet observables at RHIC.  

\section{Acknowledgments}
This material is based upon work supported financially by the National Research Foundation of South Africa and the SA-CERN Collaboration.  The author wishes to thank the Frankfurt Institute for Advanced Studies and the Nuclear Theory group at Brookhaven National Laboratory for their hospitality.

\bibliographystyle{elsarticle-num}
\bibliography{HGrefs}
\end{document}